\documentclass[twocolumn,showpacs]{revtex4}

\usepackage{graphicx}%
\usepackage{dcolumn}
\usepackage{amsmath}

\makeatletter
\def\btt#1{\texttt{\@backslashchar#1}}%
\DeclareRobustCommand\bblash{\btt{\@backslashchar}}%
\makeatother

\begin{document}


\title{Absence of magic fillings of carrier-doped C$_{60}$ in a field-effect 
transistor}

\author{Seiji Yunoki$^*$ and George A. Sawatzky$^*\dagger$}

\affiliation{
$^*$Solid State Physics Laboratory, Materials Science Center, \\ 
University of Groningen, Nijenborgh 4, 9747 AG Groningen, The Netherlands \\
$\dagger$Department of Physics and Astronomy, University of British Columbia, \\
6224 Agricultural Road, Vancouver, B.C. V6T 1Z1, Canada
}%

\date{\today}

\begin{abstract}
Motivated by recent experiments of carrier-doped C$_{60}$ in a field-effect transistor (FET), 
effects of spatial single particle potential variations on Mott-Hubbard (MH) 
insulators are studied theoretically. It is shown that the presence of strong random potentials 
leads to a reduced dependence of electronic properties on band fillings and to disappearance of 
the MH insulating behavior at integer fillings. 
A simple physical picture to explain this behavior is given using a notion of self-doping 
of the MH insulator. Our results have important implications on 
some of the puzzling observations of carrier-doped C$_{60}$ in the FET. 
It is  also discussed that the FET configuration with Al{$_2$}O{$_3$} dielectric provides 
an ``ideal'' system with strong disorders.  
\end{abstract}

\pacs{71.10.-w, 71.23.-k, 73.20.-r }

\maketitle


Carrier-doped C$_{60}$ has been known to become a Mott-Hubbard insulator 
(MHI) for {\it most} integer fillings of the three (five) fold degenerate lowest 
unoccupied (highest occupied) molecular orbitals~\cite{bruh,gunn}, 
giving a special {\it importance} to certain fillings of the system. 
This is a general property of a  Mott-Hubbard (MH) system, a good example of 
which is the cuprate high $T{\rm c}$'s in which an 
antiferromagnetic MHI with integer occupation is quickly converted to a 
metal and superconductor at modest changes of the band filling~\cite{imada}. 
It is generally accepted that electron and hole doped C$_{60}$ represents a 
MH system because the on-site Coulomb repulsion between electrons ($U\sim1.5$ eV) 
is considerably larger than the calculated band width 
($W\sim0.5$ eV)~\cite{bruh,gunn,lof}. 
Experimentally it is known that most integer filling 
C$_{60}$ systems except for perhaps K$_3$C$_{60}$ (see Refs~\cite{bruh,gunn,lof,george}) are in fact 
insulators contrary to what one finds from a band structure calculation. 
The explanation for this is generally sought in terms of a MH scenario. 
The recently developed technique of electron and hole doping using a field-effect 
transistor (FET) configuration~\cite{bat2} enables one  to study a rather 
wide and continuous  range of doping dependence of electronic states of C$_{60}$ with clear 
advantages over the use of chemical doping. Some recent results~\cite{bat} have 
already provided us with spectacular surprises which at least in some cases seem to contradict 
our general understanding of strongly correlated electrons; they do not show any 
indications of insulating behavior at or near  integer fillings!  
Here we propose one of the possible and we believe plausible  reasons for this behavior 
which has far reaching consequences regarding the interpretation of the physical properties of 
such field effect doped, correlated insulators.

\begin{figure}[thbp]
\includegraphics[clip=true,width=0.4\textwidth,angle=0]{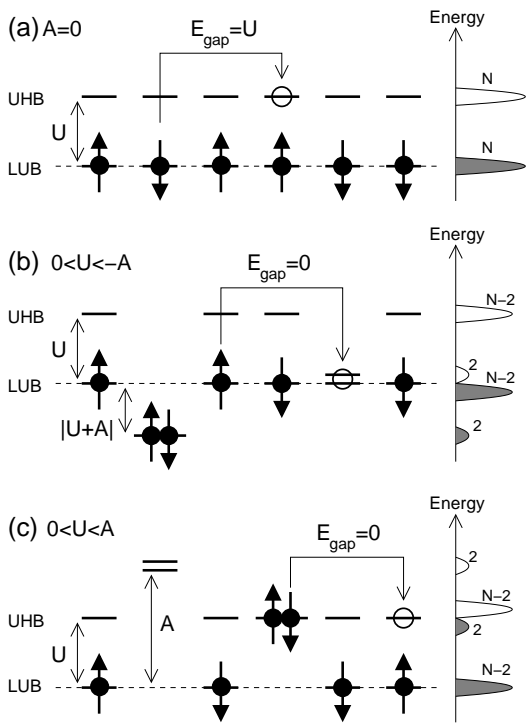}
\caption{
Schematic pictures of energy diagrams (left) and density of states (right) 
at the atomic limit for single band systems with on-site repulsions $U$ between 
electrons; (a) uniform system, (b) system with one site having a large attractive 
on-site energy $A$ ($0<U<-A$), and (c) system with one site having a large repulsive 
on-site energy ($0<U<A$). Solid circles with arrows denote electrons and their spins. 
LHB (UHB) stands for the lower (upper) Hubbard band. The numbers next to the density 
of states (left) represent the weight of the spectra with $N$ being the size of the 
system. These pictures explain $self$-$doping$ of a 
Mott-Hubbard insulator induced by a strong on-site potential energy.
}
\label{picture}
\end{figure}

The MHI is characterized by a system which, in spite of a partially filled band, has a finite 
single-particle charge excitation 
gap defined by $E_{\rm{gap}}=E_I-E_A = E_{\rm{GS}}^{N+1}+E_{\rm{GS}}^{N-1}-2E_{\rm{GS}}^{N}$ 
where $E_I=E_{\rm{GS}}^{N-1}-E_{\rm{GS}}^{N}$ ($E_A=E_{\rm{GS}}^{N}-E_{\rm{GS}}^{N+1}$) 
is the electron ionization 
(affinity) energy and 
$E_{\rm{GS}}^{N}$ the ground state energy of the system with $N$ electrons as shown in 
Fig.~\ref{picture}(a) for a non degenerate band. 
As demonstrated such a system also is characterized by local 
magnetic moments which may order below some transition temperature. It is also clear from 
Fig.~\ref{picture}(a) that this gapped situation for charge excitations only occurs for integer 
filling since both holes in the lower Hubbard band and electrons in the upper Hubbard band can move 
freely.
In Fig.~\ref{picture}(a) we see that the lower and upper Hubbard bands are well-separated 
in energy by $U$ which is the on-site coulomb repulsion of two electrons on the same site.

The experimental observations in Ref~\cite{bat} however suggest that nowhere in the electron or hole 
concentration regime in C$_{60}$ in the field effect set up does such a gap in the minimum charge 
excitation spectrum occur, since the resistivity versus gate voltage dependence is perfectly 
smooth with no signs of strongly insulating regions. We believe that this could be a result of a 
non uniform potential at the C$_{60}$-dielectric interface which led us to study the behavior of a 
MHI with strong spatial single particle potential variations.

In simple metals we expect a strong random potentials to increase the resistivity 
and perhaps even lead to localization of charge carriers~\cite{anderson} and 
insulating behavior. So it may seem strange at first glance that in a 
strongly correlated MHI strong random potentials can lead to a decrease in the 
resistivity and turn an insulator into a metal. To understand how this happens 
we look at a simple model consisting of a MHI with random on-site potentials 
added as external fields. As will be discussed later, we believe that this is close to what 
actually occurs in the C$_{60}$-dielectric interfaces in the field effect devices.

In Fig.~\ref{picture}(b)(Fig.~\ref{picture}(c)) we demonstrate what happens 
if one site for example has a large enough attractive (repulsive) on-site 
energy to overcome $U$. The physics is simple; this site becomes doubly 
occupied but the electron (hole) must come from a singly occupied site. 
Therefore aside from this site the remaining system behaves like a doped 
MHI which we know has a small or zero energy gap for charge excitations. 

\begin{figure}[thbp]
\includegraphics[clip=true,width=0.39\textwidth,angle=0]{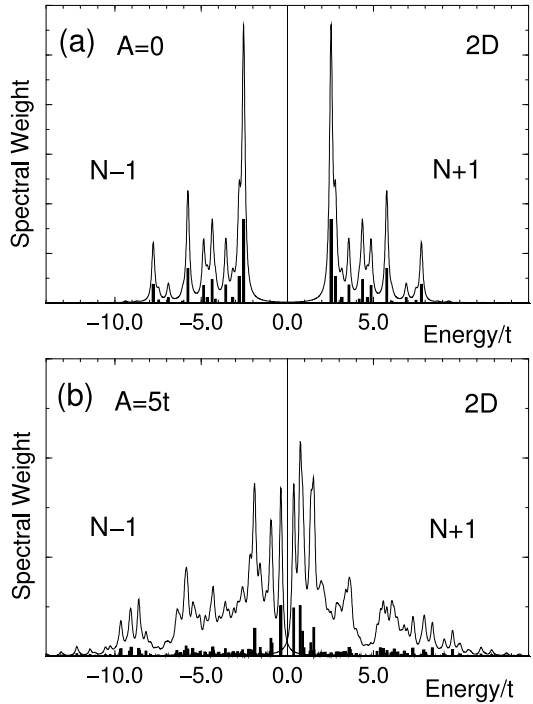}
\caption{
(a) Density of states calculated exactly for a single band 2D Hubbard cluster 
with $\sqrt{10}\times\sqrt{10}$ sites and $U=8t$ at electron density $n=1.0$. 
Solid (dashed) lines are for one-particle additional (removal) spectra using 
width of $0.1t$ to broaden $\delta$-function peaks (denoted by bars in the figure) 
into Lorentzian. 
(b) The same as in (a) but for the cluster with the site-dependent on-site random potential 
extending from $-A$ to $+A$, where $A=5t$ is taken. 
}
\label{spectra}
\end{figure}

In order to make this more realistic we look at some numerical results using 
exact diagonalization of finite size systems for the Hubbard model with a random 
on-site potential described by the Hamiltonian,
\begin{eqnarray}
H &=& - t \sum_{\langle{\bf i},{\bf j}\rangle}\sum_{\sigma} 
\left(c_{{\bf i}\sigma}^\dagger c_{{\bf j}\sigma}
+ {\rm H.c.}\right) \nonumber \\
&&+U\sum_{\bf i}
n_{{\bf i}\uparrow}n_{{\bf i}\downarrow} 
+ \sum_{\bf i}\varepsilon_{\bf i}
\left(n_{{\bf i}\uparrow}+n_{{\bf i}\downarrow}\right), 
\label{hb}
\end{eqnarray}
where $\langle{\bf i},{\bf j}\rangle$ denotes a pair of nearest-neighbor sites, 
$c_{{\bf i}\sigma}^\dagger$  ($c_{{\bf i}\sigma}$) is a creation (annihilation) operator 
of an electron with spin $\sigma(=\uparrow,\downarrow)$ at site {\bf i}, and 
$n_{{\bf i}\sigma}=c_{{\bf i}\sigma}^\dagger c_{{\bf i}\sigma}$~\cite{dent}. 
We take the random potential $\varepsilon_{\bf i}$ extending from $-A$ to $+A$ 
and a constant probability in between these extremes. The results are shown in 
Fig.~\ref{spectra} using a $\sqrt{10}\times\sqrt{10}$ site two-dimensional (2D) 
cluster with $U=8t$ where $t$ is the hopping integral. 
In Fig.~\ref{spectra}(a) we see the 
density of states for 10 electrons (half filled) and $A=0$ corresponding to the uniform system. 
The gap for charged excitations is about $U-W_{\rm eff}$ where $W_{\rm eff}$ is the effective band 
width of the lower and upper Hubbard bands, which is less than the one particle band width due to 
interactions with antiferromagnetic spin fluctuations, and $W_{\rm eff}\sim$ 3--$4t$ in this case. 
In  Fig.~\ref{spectra}(b) we see the result for 
a random potential with $A=5t$ so that the total spread of the random potential is $10t$. 
The gap for charged excitations is now all but closed and a metallic like situation will arise.

\begin{figure}[thbp]
\includegraphics[clip=true,width=0.34\textwidth,angle=0]{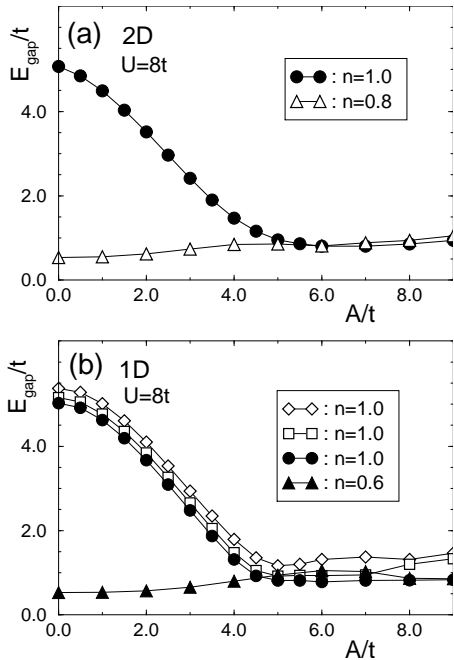}
\caption{
(a) Averaged charge gap $E_{\rm{gap}}$ sampling 100 random potential realizations for 
single band 2D Hubbard clusters with $\sqrt{10}\times\sqrt{10}$ sites and $U=8t$ 
at electron densities $n=1.0$ (solid circles) and 0.8 (open triangles). 
(b) The same as in (a) but for 1D 10-site (solid marks), 8-site (open squares), and 6-site 
chains (open diamonds). Densities are indicated in the figure. 
}
\label{gap}
\end{figure}

The variation of the gap for charged excitations with system size and with the size of the 
random potential is shown in Figs.~\ref{gap}(a) and (b). 
We see indeed that for a MHI with 
integer filling the gap rapidly decreases with increasing disorder potential saturating at a value 
close to what is expected because of the finite size nature of the calculation. 
The size dependence has been studied systematically in 1D systems (see in Fig.~\ref{gap}(b)), 
which behaves in a similar fashion in 2D. 
We also compare the gap as a function of disorder potential for non-integer filling 
systems consisting of 8 and 6 electrons in 10 sites in Figs.~\ref{gap}(a) and (b), respectively. 
This system would ideally be metallic for no disorder but 
for the finite size still shows a gap because of this and then first the gap increases with 
increasing disorder potential and then turns over and decreases for large disorder potential and 
ends up close to that of the integer filled system. This result clearly demonstrates that a MHI 
with strong disorder has a charge gap which is not or only weakly dependent on the filling 
and that integer filling is no longer a special case.

Why is this relevant for C$_{60}$ in the FET-like device measurements? We recall how the devices 
are actually made. One starts with a very pure crystal of C$_{60}$ and evaporates onto this the 
source and drain electrodes which often is gold. Then using plasma sputtering a thin layer of 
10--20 nm of Al{$_2$}O{$_3$} is sputtered onto the C$_{60}$ which serves as the dielectric 
and on top of this 
the gate metal electrode. Upon application of a gate voltage calculations show that the charge 
carriers are concentrated in the first C$_{60}$ layer at the C$_{60}$-dielectric interface. 
This means that even if the C$_{60}$ surface layer was perfectly smooth the C$_{60}$ molecules 
at the interface will in fact experience a spatially non uniform potential because the dielectric 
is not a single crystal. In fact for Al{$_2$}O{$_3$} the interface C$_{60}$ molecules will see in 
their close vicinity a variety of 
crystallite orientations and crystal faces. We now recall that Al{$_2$}O{$_3$} is an ionic insulator 
stabilized to a large extend by the Madelung potential set up by the ionic charges. It is easy to 
see that the Madelung potential fields in the vicinity of the crystallite will leak outward to 
distances of at most several Al{$_2$}O{$_3$} lattice spacings and that the strength of these fields will 
depend strongly on the crystallite orientation and termination. The potential at distances of 0.3 
nm from the interface can easily be as large as $\pm2$ eV and can also vary by that amount 
depending on the crystal orientation and termination. The Hubbard $U$ has been measured on 
the C$_{60}$ (111) surface and is found to be 1.6eV~\cite{lof} and the band width is about 0.5 eV. 
Therefore indeed the spatial variation of the potential can be expected to be of the order of the 
MH gap resulting in a strongly disordered system with a charge excitation gap which will be 
nearly independent of the charge concentration. 

This may be the answer to the question related to the absence of insulating regions in the 
regions where integer doping is expected but it also then raises a lot of other questions. For 
example why is the mobility so high and why does the charge motion look like coherent band 
motion? We do not pretend to know the answer but it is important to note that we do not need 
a large density of potentials with variations of 1 eV or so. If 10\% of the interface C$_{60}$ 
molecules experienced such potential fluctuations this is enough to wipe out the MHI regimes. 
This could still leave enough space for percolating regions of weak scattering. Another very 
interesting thought could be that strong disorder in the first layer could cause some charge 
carriers to actually move to the second layer and that the high mobility is caused by motion in 
the second layer. We note that the Madelung potential like effects decay exponentially as one 
moves away from the interface so the second layer would experience at most weak scattering 
effects. The question as to the influence on superconductivity remains an open question. Why 
would doped carriers in such disordered systems favor a superconducting state rather than a 
glassy-like state which is likely the case in ``dirty'' boson systems~\cite{fisher}.

To summarize, we have studied the effects of on-site potential variations on a MHI 
and shown that strong disorder leads to a reduced dependence of electronic properties 
on band fillings and to disappearance of the MH insulating behavior at integer fillings. 
We also discussed that carrier-doped C$_{60}$ in the FET configuration 
with Al{$_2$}O{$_3$} dielectric may provide an ideal situation for a strongly disordered 
MH system. Our results explain some of the most puzzling experimental observations for 
carrier-doped C$_{60}$ in the FET.





\vskip-0.5cm
\begin{references}


\bibitem{bruh} P. Rudolf, M. S. Golden, and P. A. Br\"uhwiler, 
J. Electr. Spectros. Relat. Phenom. {\bf 100}, 409 (1999). 

\bibitem{gunn} O. Gnunnarsson,  
Rev. Mod. Phys. {\bf 69}, 575 (1997). 

\bibitem{imada} See {\it e.g.}, M. Imada, A. Fujimori, and Y. Tokura, 
Rev. Mod. Phys. {\bf 70}, 1039 (1998). 



\bibitem{lof} R. W. Lof, M. A. van Veenendaal, B. Koopmans, H. T. Jonkman, and 
G. A. Sawatzky, Phys. Rev. Lett. {\bf 68}, 3924 (1992).

\bibitem{george} G. A. Sawatzky, in {\it Physics and Chemistry of Fullerenes 
and Derivatives}, edited by H. Kuzmany, J. Fink, M. Mehring, and S. Roth, 
(World Scientific, Singapore, 1995), p. 373. 



\bibitem{bat2} J. H. Sch\"on, Ch. Kloc, R. C. Haddon, and B. Batlogg, 
Science {\bf 288}, 656 (2000).


\bibitem{bat} J. H. Sch\"on, Ch. Kloc, and B. Batlogg, 
Nature {\bf 408}, 549 (2000).



\bibitem{anderson} P. W. Anderson, Phys. Rew. {\bf 109}, 1492 (1958). 


\bibitem{dent} Finite temperature properties of similar models have been reported recently by 
P. J. H. Denteneer, R. T. Scalettar, and N. Trivedi in Phys. Rev. Lett. {\bf 87}, 146401 (2001) 
with a special emphasis on the symmetry possessed by the models. 



\bibitem{fisher} M. P. A. Fisher, P. B. Weichman, G. Grinstein, and 
D. S. Fisher, Phys. Rev. B {\bf 40}, 546 (1989).

\end{references}
\end{document}